\documentclass[11pt]{article}
\usepackage{amsmath}
\usepackage{graphicx}
\usepackage{ textcomp }
\usepackage{ amssymb }
\usepackage{float}
\graphicspath{ {images/} }
\usepackage{booktabs} 
\usepackage[table]{xcolor} 
\usepackage{authblk}

\usepackage{tabularx} 
\usepackage{rotating} 
\usepackage{siunitx} 
\usepackage{caption}

\usepackage{hyperref}
\usepackage[numbers,sort&compress]{natbib}

\usepackage{adjustbox}
\usepackage[margin=0.5in]{geometry}
\begin{document}

\makeatletter
\newcommand*{\rom}[1]{\expandafter\@slowromancap\romannumeral #1@}
\makeatother

\title{Pseudo-perplexity in One Fell Swoop for 
       Protein Fitness Estimation}

\author{
  Pranav Kantroo\textsuperscript{1,6,}\footnote{pranav.kantroo@yale.edu} ,
  G\"unter P. Wagner\textsuperscript{2,3,4}, 
  Benjamin B. Machta\textsuperscript{5,6,}\footnote{benjamin.machta@yale.edu}
  \\
  \small
  \textsuperscript{1}Computational Biology and Bioinformatics Program, Yale University, New Haven, CT-06520, USA \\
  \textsuperscript{2}Emeritus, Department of Ecology and Evolutionary Biology, Yale University, New Haven, CT-06520, USA \\
  \textsuperscript{3}Department of Evolutionary Biology, University of Vienna, Djerassi Platz 1, A-1030 Vienna, Austria \\
  \textsuperscript{4}Hagler Institute for Advanced Studies, Texas A\&M, College Station, TX-77843, USA \\
  \textsuperscript{5}Department of Physics, Yale University, New Haven, CT-06520, USA \\
  \textsuperscript{6}Quantitative Biology Institute, Yale University, New Haven, CT-06520, USA 

}
\date{}
\maketitle

\begin{abstract}
\noindent Protein language models trained on the masked language modeling objective learn to predict the identity of hidden amino acid residues within a sequence using the remaining observable sequence as context. They do so by embedding the residues into a high dimensional space that encapsulates the relevant contextual cues. These embedding vectors serve as an informative context-sensitive representation that not only aids with the defined training objective, but can also be used for other tasks by downstream models. We propose a scheme to use the embeddings of an unmasked sequence to estimate the corresponding masked probability vectors for all the positions in a single forward pass through the language model. This One Fell Swoop (OFS) approach allows us to efficiently estimate the pseudo-perplexity of the sequence, a measure of the model’s uncertainty in its predictions, that can also serve as a fitness estimate. We find that ESM2 OFS pseudo-perplexity performs nearly as well as the true pseudo-perplexity at fitness estimation, and more notably it defines a new state of the art on the ProteinGym Indels benchmark. The strong performance of the fitness measure prompted us to investigate if it could be used to detect the elevated stability reported in reconstructed ancestral sequences. We find that this measure ranks ancestral reconstructions as more fit than extant sequences. Finally, we show that the computational efficiency of the technique allows for the use of Monte Carlo methods that can rapidly explore functional sequence space.

\end{abstract}

\section{Introduction}

The field of protein engineering is in the midst of a renaissance. Deep learning-based methods have enabled unprecedented strides towards addressing questions relating to the folding, design and fitness prediction of protein sequences \citep{jumper2021highly, dauparas2022robust, watson2023novo, notin2022tranception}. Large language models based on the transformer architecture lie at the heart of powering a host of such techniques \citep{vaswani2017attention}. Protein language models are typically composed of stacks of such transformer blocks, that are trained on data derived from large swaths of protein sequences to model the underlying data distribution \citep{lin2022language, nijkamp2023progen2}. The training data can take the form of raw sequences, alignments of related sequences, protein structures and any combination of these modalities \citep{lin2022language, rao2021msa, hsu2022learning, heinzinger2023prostt5}. 

Just as sentences are different from random strings of words, proteins too are different from random strings of amino acid residues. Proteins being products of evolution, adhere to what has been termed \textit{Protein Grammar}- a conceptual anchor that connotes the rules governing the underlying generative process~\citep{verkuil2022language}. Our ability to design effective proteins hinges on being able to parse and characterize the dependencies that define this grammar. The sheer scale of the data, combined with the enormous number of tunable parameters, allows machine learning models to implicitly learn these operational principles \citep{bepler2021learning}. For models that exclusively utilize raw sequences, two broad modeling paradigms have emerged towards this end: the masked language modeling objective and the next-token prediction objective~\citep{ruffolo2024designing}. 

The Masked language modeling objective involves learning to predict the identity of hidden amino acid residues in a sequence using the remaining observable sequence as context \citep{devlin2018bert, lin2022language}. The objective forces the model to assimilate both general and context-specific features within the input sequence to identify which residues fit best at the masked positions. A model trained on this objective is called an encoder model, since it transforms the residues in the input sequence into a set of high dimensional vectors that encapsulates its understanding of the given input. This set of high dimensional vectors is called the embedding, which is a task-agnostic, informative representation of the sequence that can be leveraged by downstream models to perform specific tasks, like protein folding \citep{alley2019unified, elnaggar2021prottrans, lin2022language}. 

The Next-token prediction objective, on the other hand requires learning to predict the identity of the next amino acid residue using the preceding sequence as context \citep{radford2018improving}. Models trained on this objective are called decoder models or autoregressive models, since they can be used to generate novel sequences by iteratively sampling the next amino acid in a scheme where the output of the model at each step is fed back as the input for the next step \citep{shin2021protein, notin2022tranception, ferruz2022protgpt2, hesslow2022rita}. The autoregressive sampling scheme illustrates how the nature of the training objective shapes the use-cases that these models can be applied to: decoding models are naturally well-suited for generative tasks in protein design.

Both these classes of models can be used to quantify the degree to which an input sequence adheres to the rules of protein grammar as learned by the models. This is accomplished through the use of various heuristics that transform the raw model output- a set of probability distributions on the amino acid alphabet, into a scalar quantity. Such scalar estimates that measure adherence to protein grammar have been empirically validated to be related to protein fitness: the more a sequence conforms to the rules of the learned grammar, the higher its fitness tends to be. As a consequence, language models have found extensive use in variant effect prediction, which entails assessing how mutations in a sequence affect its function~\cite{horne2022recent, meier2021language, notin2022tranception, notin2024proteingym}. 

Decoder models have excelled at this fitness estimation task due to their ability to score protein variants with all three kinds of mutations: substitutions, insertions and deletions. Encoder models however have been limited in their scope to evaluating variants that differ only by substitutions~\citep{notin2022tranception, notin2024proteingym}. The reason for this discrepancy can be attributed to the differences in the properties of the respective heuristics that are used to arrive at the fitness assessments. In this manuscript we develop a simple extension that enables encoders to be used for fitness estimation of indels, and also allows for their wider use for generative tasks in protein design.

\begin{figure}[htbp]
    \centering
    \includegraphics[width=0.9\textwidth]{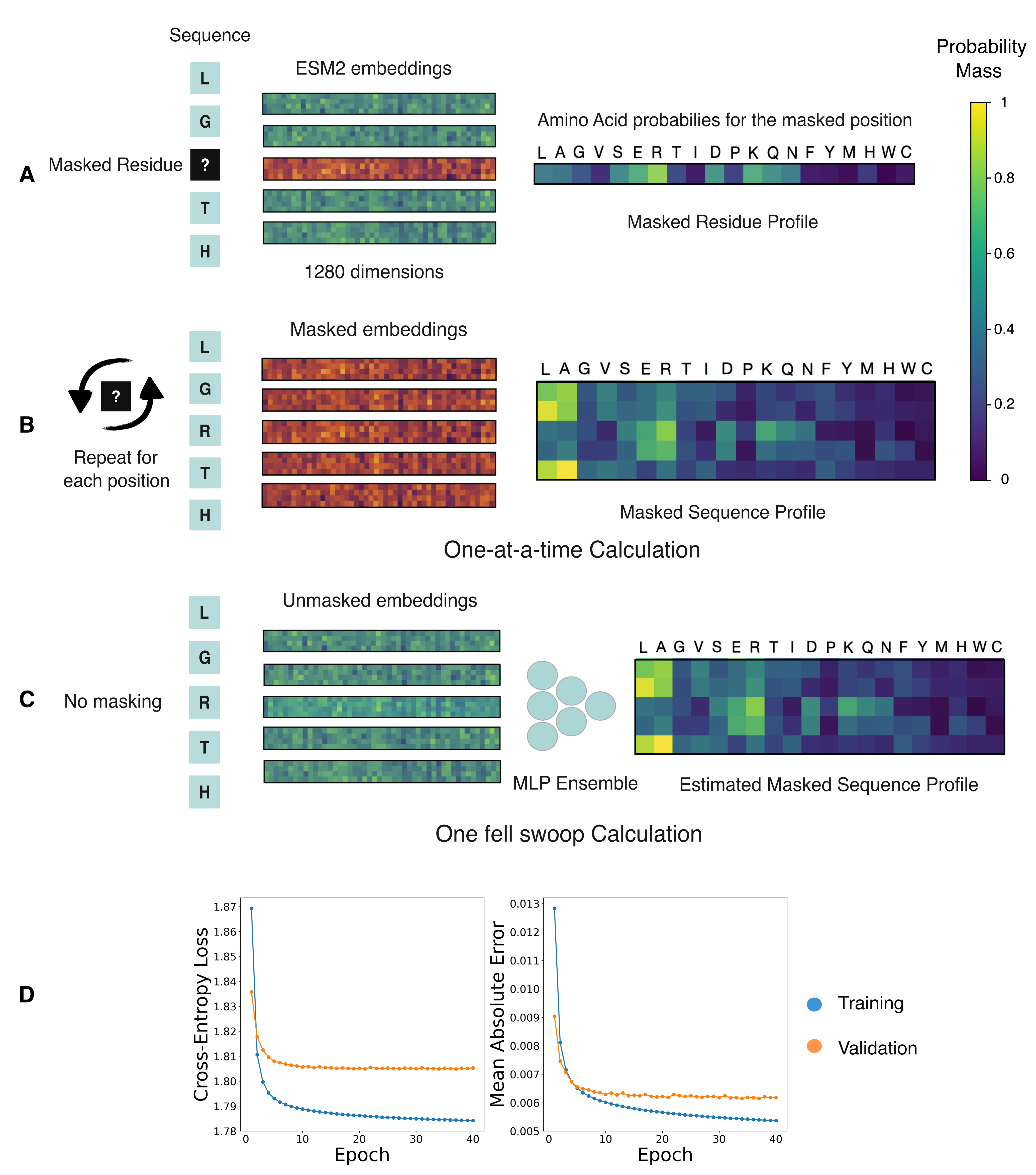}
    \caption{\small (A) The masked residue profile for the sequence where the mask is placed at position 3. The embeddings generated by the model take the form of a high dimensional vector, 1280 dimensional in this case. The embeddings for the unmasked residues are colored in green, while the embedding for the masked residue is in orange. 
    (B) A representation of the masked sequence profile calculation for the entire sequence where each residue is masked one-at-a-time to generate the corresponding masked residue profile. The sequence profile is obtained by collating the individual residue profiles into a matrix. 
    (C) The One Fell Swoop setup, where the unmasked residue embeddings are used to estimate the corresponding masked residue profile for the position using an ensemble of multi-layer perceptrons. The network is trained to minimize the cross-entropy loss between its prediction and the output of the language model.
    (D) The loss curves of the One Fell Swoop network on the training and validation splits of the dataset. The model was trained to minimize the cross-entropy loss. The mean absolute error is used to show the average deviation in the individual elements of the sequence profile matrix. Each element in the matrix corresponds to a probability, and takes values between 0 and 1.}
    \label{fig:figure_1}
\end{figure}

\section{Quantifying adherence to Protein Grammar}

In encoder models, the prediction for a specific masked position takes the form of a probability vector, where the dimensions of the vector denote the respective tokens in the model’s vocabulary- amino acid residues in this case. The values in this vector represent the learned estimates for the probability that a given residue will stand in place of the mask token, conditioned on the context from the unmasked sequence (Figure 1:A). In this setting, the magnitude of the probability mass for a residue can be used as a proxy for the degree to which the residue fits in the sequence at that specific position. Consequently, residues with a higher probability mass should map to sequences that are more likely to be functional, and therefore have a higher fitness score. This intuition is formalized via the masked marginal heuristic to evaluate mutations \citep{meier2021language}. 

Let $M$ be the set of positions in a wild-type sequence $S^\mathrm{wt}$ where substitutions have been introduced to create a mutant sequence $S^\mathrm{mt}$. The masked marginal (MM) heuristic dictates that the fitness $F$ of the mutant sequence relative to the wild-type sequence can be scored as follows:
\begin{equation}
F_\mathrm{MM}(S^\mathrm{mt}|S^\mathrm{wt}) = \sum_{i \in M} \log \frac{P(r_{i} = r_{i}^\mathrm{mt}|S_{-M})}{P(r_{i} = r_{i}^\mathrm{wt}|S_{-M})}
\end{equation}
where $r_i$ is used to denote the $i^{th}$ amino acid residue in a sequence and $S_{-M}$ is the protein sequence masked at all the positions where the substitutions have been introduced. A positive score means that the mutant has a higher fitness than the wild-type sequence, and a negative score implies that the opposite is true. Note that this evaluation heuristic is based on a log-likelihood ratio, and is therefore by definition limited in its use to scoring sequences that differ by substitutions. This is because residues that have been inserted or deleted have no equivalent counterparts to compare against in this evaluation scheme.

Decoder models circumvent this issue by evaluating sequences using an altogether different heuristic. The heuristic relies on the idea that the fitness of a sequence is proportional to its likelihood as a whole, as measured by the language model: the more typical a sequence looks, the higher its fitness score should be \citep{notin2022tranception, ruffolo2024designing}. The likelihood of a sequence $S$ of length $L$ can be expressed as a product of terms where each term only depends on the preceding context in the sequence:
\begin{equation}
P(S) = \prod_{i=1}^{L} P(r_{i}=r_{i}^{S}|S_{<i})
\end{equation}
where $S_{<i}$ denotes the identity of the residues in the sequence up until the $(i-1)^{th}$ position. 
Decoder models can efficiently compute each term in the factorized expression in a parallelized fashion, yielding the individual terms to calculate the log-likelihood of the sequence. Normalizing the log-likelihood by the sequence length enables fitness comparisons across sequences with different lengths, elucidating why the heuristic also works for indel mutations~\citep{madani2023large}. The fitness evaluation metric used in decoder models thus typically takes the form of the per-residue log-likelihood (LL) that can be directly used for fitness comparisons across closely related sequences in a reference-free manner: the higher the per-residue log-likelihood, the higher the fitness of the sequence.
\begin{equation}
F_\mathrm{LL}(S) = \frac{1}{L} \sum_{i=1}^{L} \log(P(r_{i}=r_{i}^{S}|S_{<i}))
\end{equation}

Decoder models, by virtue of their ability to score both substitutions and indels via the per residue sequence log-likelihoods, have thus become the de facto standard for the fitness estimation task \cite{notin2024proteingym, ruffolo2024designing}. However, an equivalent analog in the form of a pseudo-log-likelihood, also exists for encoder models \cite{salazar2019masked}. The per-residue pseudo-log-likelihood (PLL) of a sequence, measured via an encoder model can be expressed as follows:

\begin{equation}
F_\mathrm{PLL}(S) = \frac{1}{L} \sum_{i=1}^{L} \log(P(r_{i}=r_{i}^{S}|S_{-i}))
\end{equation}
where $S_{-i}$ denotes the sequence $S$ masked at the position $i$. 

In principle, the pseudo-log-likelihoods obtained from encoder models could be used as a fitness measure for indel mutations. We contend that the primary deterrent for their use in this task is that the expression is prohibitively slow to compute. This is because unlike decoder models, each of the $L$ terms in the expression needs to be calculated one-at-a-time by introducing the mask token at each position in the sequence. The pseudo-log-likelihood computation thus consumes $L$ forward passes through the model for a sequence of length $L$ (Figure 1:B)~\cite{salazar2019masked}. In comparison, the masked-marginal strategy and the decoder-based parallelized log-likelihood computation, both take just one forward pass regardless of sequence length.

\section{Pseudo-perplexity in One fell swoop}

We get around the length dependency of the pseudo-log-likelihood calculation by proposing a One Fell Swoop (OFS) approach that can make do with just a single forward pass through the encoder model. The approach relies on the idea that the embedding of an unmasked residue contains enough information to be able to accurately infer the residue's masked profile. If this is true, then it should be possible to obtain the masked residue profile of the  $i^\text{th}$ amino acid, $P(r_{i}=r_{i}^{S}|S_{-i})$, by projecting it from its unmasked embedding vector $E_i$ using some learned function~$f$:

\begin{equation}
f(E_i) \sim P(r_{i}=r_{i}^{S}|S_{-i})
\end{equation}

Such a projection function from the embedding space to the space of probability vectors, should enable single pass parallelized estimation of the pseudo-log-likelihood of the sequence. We thus reframe the pseudo-log-likelihood calculation as a regression problem to estimate the entire sequence profile using its unmasked residue embeddings. The function takes the form of a neural network model- an ensemble of multi-layer perceptrons, and the training data to learn the mapping is generated by performing the one-at-a-time calculation for a large corpus of sequences. The sequences used in this step are the representative sequences for protein clusters obtained by clustering a protein sequence database via mmseqs2 \cite{steinegger2017mmseqs2}. The model is trained to minimize the cross-entropy loss between its prediction and the profile matrix obtained via the one-at-a-time calculation (Figure 1:B and Figure 1:C).

As a proof of concept, we train a model to generate the ESM2 (650M) sequence profile of an input sequence using its ESM2 (650M) unmasked residue embeddings \cite{lin2022language}. The loss curves for the training and validation data splits over the course of the training are shown in Figure 1:D. We observe that the learned projection from the embedding space to the sequence profile substantially improves over the course of the training. The mean absolute error quantifies the error in the individual elements of the sequence profile matrix. The small magnitude of the mean absolute error in the prediction of the trained model indicates that the procedure can be effectively used to approximate the sequence profile. 

The method as outlined above can be used for any encoder model, however its effectiveness is contingent on the quality of the embeddings being used, and the relative difficulty of the regression task.  The reason for the boost in efficiency can be pinned to the fact that the unmasked embeddings can be calculated in a single forward pass through the encoder model, and the subsequent profile calculation is performed in parallel by the much tinier one fell swoop setup. Calculating an estimate of the pseudo-log-likelihoods of the sequence can thus be turned into a single pass affair, while also generating the entire single residue substitution-based mutational landscape of the sequence in the process.

In practice, we use the sequence profile obtained from the one fell swoop procedure to calculate the pseudo-perplexity (PP) $Z$, for fitness estimation. It is a quantity that is derived from the per-residue pseudo-log-likelihood and can be expressed as \cite{salazar2019masked}:

\begin{equation}
Z(S) = e^{-F_\mathrm{PLL}(S)} 
\end{equation}

Pseudo-perplexity measures the uncertainty in an encoding model's predictions for some given input. A low pseudo-perplexity indicates that the input sequence is in line with the assessment of the model based on what it learnt from the data that it was trained on. It therefore represents how natural or typical the sequence seems to the model: lower the pseudo-perplexity, more natural the sequence looks. Pseudo-perplexity is inversely related to fitness, and its negative value can serve as a fitness estimate. Although exponentiating the per-residue pseudo-log-likelihood does not add utility, it does enhance interpretability and is usually a small number for natural protein sequences. Note that the One fell swoop procedure produces an estimate of the ESM2 (650M) sequence profile, and not its exact value. Pseudo-perplexity calculated using this estimated sequence profile is denoted by OFS pseudo-perplexity to draw the distinction.

\section{Evaluating OFS pseudo-perplexity as a fitness estimator}

The reliability of computational fitness estimates can be assessed by comparing them against experimentally measured fitness scores from Deep Mutational Scanning (DMS) experiments. DMS as a technique has made it possible to conduct large-scale probes into how mutations affect protein function. This has paved the way to chart the localized genotype-to-phenotype map for several proteins of interest. Broadly, such an experiment entails introducing random mutations into a specific gene for an entire cell population. The fitness of the respective variants in the population can then be estimated through the changes in their relative abundance over the course of a selection assay. The underlying idea here is that the high fitness variants will be enriched in the population, while dysfunctional variants will end up being depleted in this process  \citep{fowler2014deep}. The degree of concordance between the language model fitness estimates and the experimentally determined ground truth can thus be used as the performance metric to assess how well these models do on the zero-shot fitness estimation task. 

The ProteinGym benchmark is a compilation of data obtained from a large set of DMS experiments that spans a wide variety of selection assays: Activity, Binding, Expression, Organismal Fitness and Stability. The extensive dataset can thus be used to benchmark the performance of these models and systematically document methodological improvements \citep{notin2024proteingym}. We tested the efficacy of ESM2 OFS pseudo-perplexity at zero-shot fitness prediction using experimentally determined fitness measurements drawn from DMS experiments and compared its performance with other baselines using data sourced from the ProteinGym repository. The consistency between the experimentally measured fitness scores and the language model assessment is measured using the Spearman rank correlation. The metric measures the correlation between the rank orderings of the fitness scores between the two sets.

%\captionsetup[table]{font=bf}
\captionsetup[table]{}
\newcolumntype{C}[1]{>{\centering\arraybackslash}m{#1}}
\newcolumntype{L}[1]{>{\raggedright\arraybackslash}m{#1}}

\begin{table*}[ht]
\centering

\begin{tabular}{@{}L{4cm}C{2cm}C{2cm}C{2cm}C{2cm}C{2cm}C{2cm}@{}}
\toprule
& \multicolumn{5}{c}{\textbf{Function Type}} & \\
\cmidrule(lr){2-6}
\textbf{Language Model} & \textbf{Activity} & \textbf{Binding} & \textbf{Expression} & \textbf{Organismal Fitness} & \textbf{Stability} & \textbf{Aggregate Mean}\\
& (43) & (14) & (17) & (77) & (66) &  \\
\midrule
ESM2: MM& 0.430 & 0.332 & 0.440 & 0.363 & 0.518 & 0.430 \\
ESM2: OFS PP & 0.393 & 0.279 & 0.397 & 0.331 & 0.507 & 0.403 \\
\midrule
CARP (640M) & 0.395 & 0.274 & 0.419 & 0.364 & 0.414 & 0.385 \\
ESM-1v (single) & 0.390 & 0.268 & 0.431 & 0.362 & 0.476 & 0.405 \\
Progen2 M & 0.393 & 0.296 & 0.436 & 0.381 & 0.396 & 0.388 \\
RITA L & 0.359 & 0.291 & 0.422 & 0.374 & 0.383 & 0.375 \\
Tranception L no retrieval & 0.401 & 0.289 & 0.415 & 0.389 & 0.381 & 0.386 \\

\bottomrule
\end{tabular}

\caption{ProteinGym Substitutions: Spearman correlation}

\end{table*}

\begin{table*}[ht]
\centering

\begin{tabular}{@{}L{4cm}C{2cm}C{2cm}C{2cm}C{2cm}C{2cm}@{}}
\toprule
& \multicolumn{4}{c}{\textbf{Function Type}} & \\
\cmidrule(lr){2-5}
\textbf{Language Model} & \textbf{Activity} & \textbf{Expression} & \textbf{Organismal Fitness} & \textbf{Stability} & \textbf{Aggregate Mean}\\
& (3) & (2) & (6) & (55) & \\
\midrule
ESM2: OFS PP & 0.518 & 0.433 & 0.553 & 0.582 & 0.574 \\
\midrule
Progen2 M & 0.510 & 0.466 & 0.374 & 0.517 & 0.503 \\
RITA L & 0.466 & 0.456 & 0.419 & 0.495 & 0.486 \\
Tranception L no retrieval & 0.431 & 0.412 & 0.445 & 0.469 & 0.464 \\

\bottomrule
\end{tabular}

\caption{ProteinGym Indels: Spearman correlation}

The data for the baselines (CARP, ESM-1v, Progen2 M, RITA L and Tranception L no retrieval) is sourced from the ProteinGym repository.

\end{table*}

We assessed the reliability of the OFS pseudo-perplexity fitness estimate by comparing its performance against pseudo-perplexity obtained via the one-at-a-time calculation for 61 DMS assays. We observe that the performance of the two measures is largely consistent, with true pseudo-perplexity having a slight edge over OFS pseudo-perplexity (Figure 2:A). We also found that OFS pseudo-perplexity as an evaluation strategy incurs a modest decline in performance on the substitutions benchmark, as compared to the masked marginal heuristic (Figure 2:B and Table 1). This observation is consistent with previously reported results that assessed the performance of pseudo-perplexity \citep{meier2021language}, and shows that the masked marginal heuristic is the preferable evaluation strategy for ESM2 when exclusively dealing with substitutions. Comparison with other baselines indicates that ESM2 OFS pseudo-perplexity as a stand-alone fitness estimator is competitive with state of the art protein language models in its size class that rely exclusively on sequence information (Figure 2:B and Table 1). The utility of the approach at fitness estimation is more apparent on the Indels benchmark where encoding models have not been tested thus far. Not only does OFS allow encoding models like ESM2 to be used as fitness estimators for insertions and deletions, we find that ESM2 OFS pseudo-perplexity defines a new state of the art on the Indels benchmark (Figure 2:C and Table 2).

\begin{figure}[htbp]
    \centering
    \includegraphics[width=0.84\textwidth]{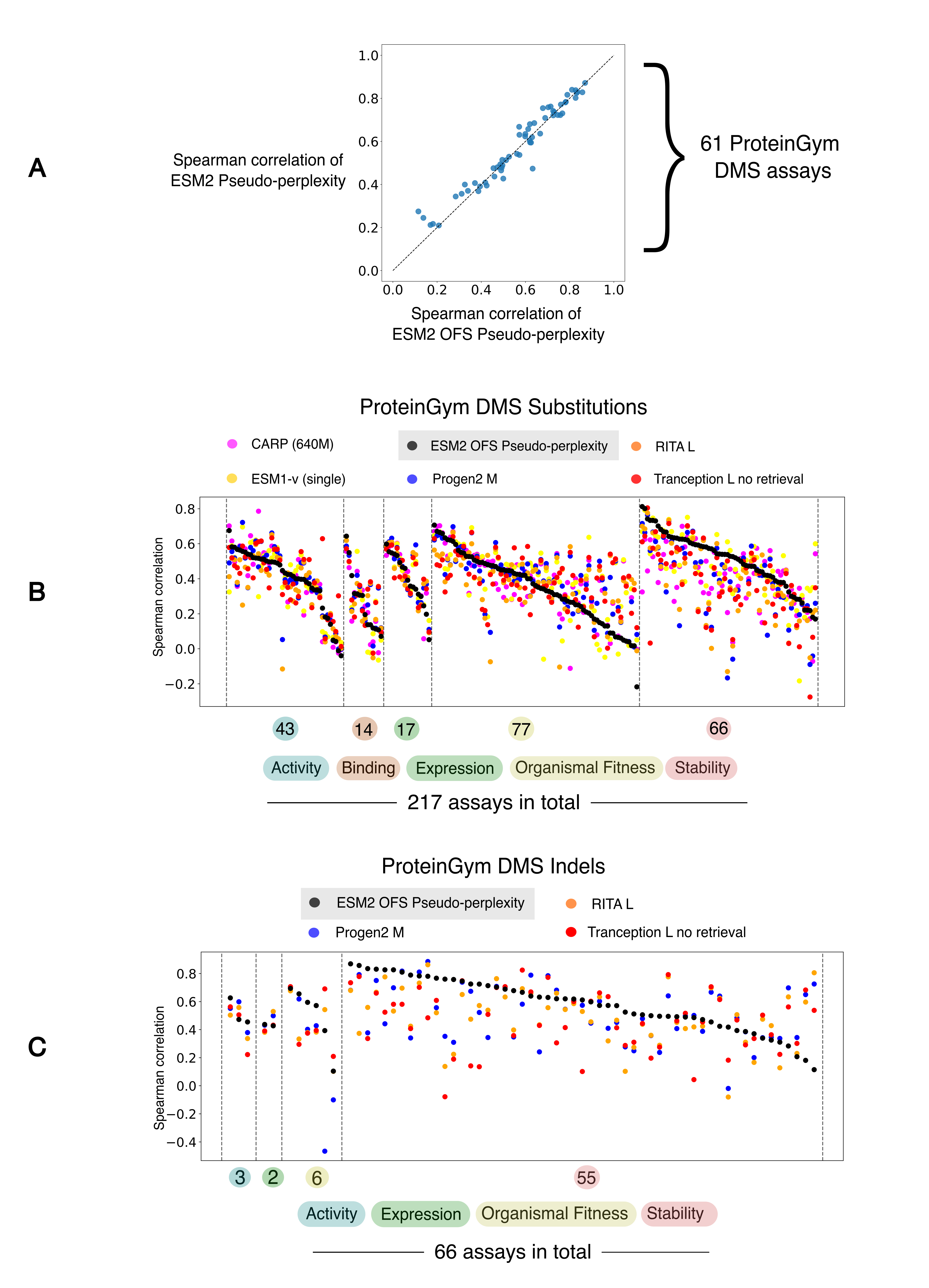}
    \caption{\small (A) The performance of the One fell swoop calculation at the fitness estimation task is            similar to the performance of the one-at-a-time calculation. 
            (B) The benchmarking data is arranged by function types, and the individual DMS experiments within each category are arranged in descending order of the performance of ESM2 OFS pseudo-perplexity as measured by the Spearman correlation. ESM2 OFS pseudo-perplexity is competitive with other state-of-the-art protein language models in its size class that exclusively utilize sequence information. The performance data for the other models was obtained from the ProteinGym repository.
            (C) ESM2 OFS pseudo-perplexity defines a new state of the art on the ProteinGym Indels benchmark.}
    \label{fig:figure_2}
\end{figure}

We note that we have chosen to compare ESM2 only against similarly-sized models that have been trained exclusively on raw protein sequences. Hybrid models can utilize additional data modalities like multiple sequence alignments or structure, and are known to exhibit significant overall improvements in their benchmark performance compared to models that are trained exclusively on sequences \citep{notin2022tranception, rao2021msa, su2023saprot, li2024prosst, paul2023combining}. The trend makes intuitive sense because sequence alignments and structure are richer modalities that explicitly contain information about the evolutionary context and the three dimensional configuration that the protein acquires in space. This additional information in the form of evolutionarily constrained and flexible positions, the empirical distribution of viable residues at a specific position, or the physical constraints imposed due to structure, can allow the model to identify broader trends and refine its assessment. 

Notably, it has also been shown that models trained on specific data modalities can excel at capturing particular aspects of protein fitness. Specifically, inverse-folding models, that predict sequence likelihoods by using the structure of the sequence as the conditioning variable, along with hybrid models that incorporate structural information, have been found to be especially adept at fitness estimation for stability-based assays within ProteinGym \citep{paul2023combining, su2023saprot, li2024prosst}. Data drawn from specific selection assays can thus be used to conceptualize the model assessment in terms of the aspect of protein fitness that it best correlates with. In this context, we note that ESM2 OFS pseudo-perplexity is most strongly correlated with data drawn from stability-based assays, as is evident from both the substitutions and the indels benchmark (Table 1 and Table 2). This observation led us to our next question: Can the fitness measure be used to capture the widely-reported elevated stability of reconstructed ancestral sequences?~\citep{trudeau2016potential}

\section{Ancestral state reconstruction: Old is Gold?}

Sequence alignments of related extant sequences can be used to infer their phylogenetic relationship, and study the functional evolution of the family over time through the topology of the inferred tree \citep{kapli2020phylogenetic}. The topology along with the alignment data, can in turn be used to estimate the states of the ancestral nodes in the tree through various statistical frameworks, in a process called ancestral state reconstruction \citep{joy2016ancestral}. The technique has made it possible to resurrect proteins that may have existed eons ago, and thereby directly investigate the functional properties of these inferred ancestral sequences  \citep{hochberg2017reconstructing, pillai2020origin, field2006adaptive, lynch2011regulatory, nnamani2016derived, brayer2011evolution}. The research program has led to the rather perplexing finding that reconstructed ancestral sequences tend to have superior functional properties compared to their extant counterparts: improved activity, enhanced stability and/or higher promiscuity \citep{hendrikse2021ancestral, gaucher2008palaeotemperature, risso2013hyperstability, risso2014thermostable}. This is true, so much so that, ancestral state reconstruction has emerged as a viable strategy for protein design \citep{spence2021ancestral, johnson2024computational}.

The phenomenon as a whole is not well understood. One line of reasoning to explain the trend is that the technique suffers from an inherent bias in the form of a consensus effect, that systematically leads to sequences with a higher stability. The idea is that if a given residue is predominantly found in a specific position in a sequence alignment, then it is likely to be assigned as the ancestral state for the reconstructed ancestors. This reasoning is well-founded for positions that are constrained by function, however, it may also lead to spurious inference that is biased towards the consensus for positions where this is not true~\citep{trudeau2016potential}. Replacing rare residues by the consensus residues drawn from the sequence alignment has also been shown to generate sequences with higher stability \citep{pey2008engineering, sullivan2012stabilizing, komor2012highly, sternke2019consensus}. In this view, the mystical properties of ancestral sequences can be traced back to this bias towards the consensus. 

\begin{figure}[htbp]
    \centering
    \includegraphics[width=0.85\textwidth]{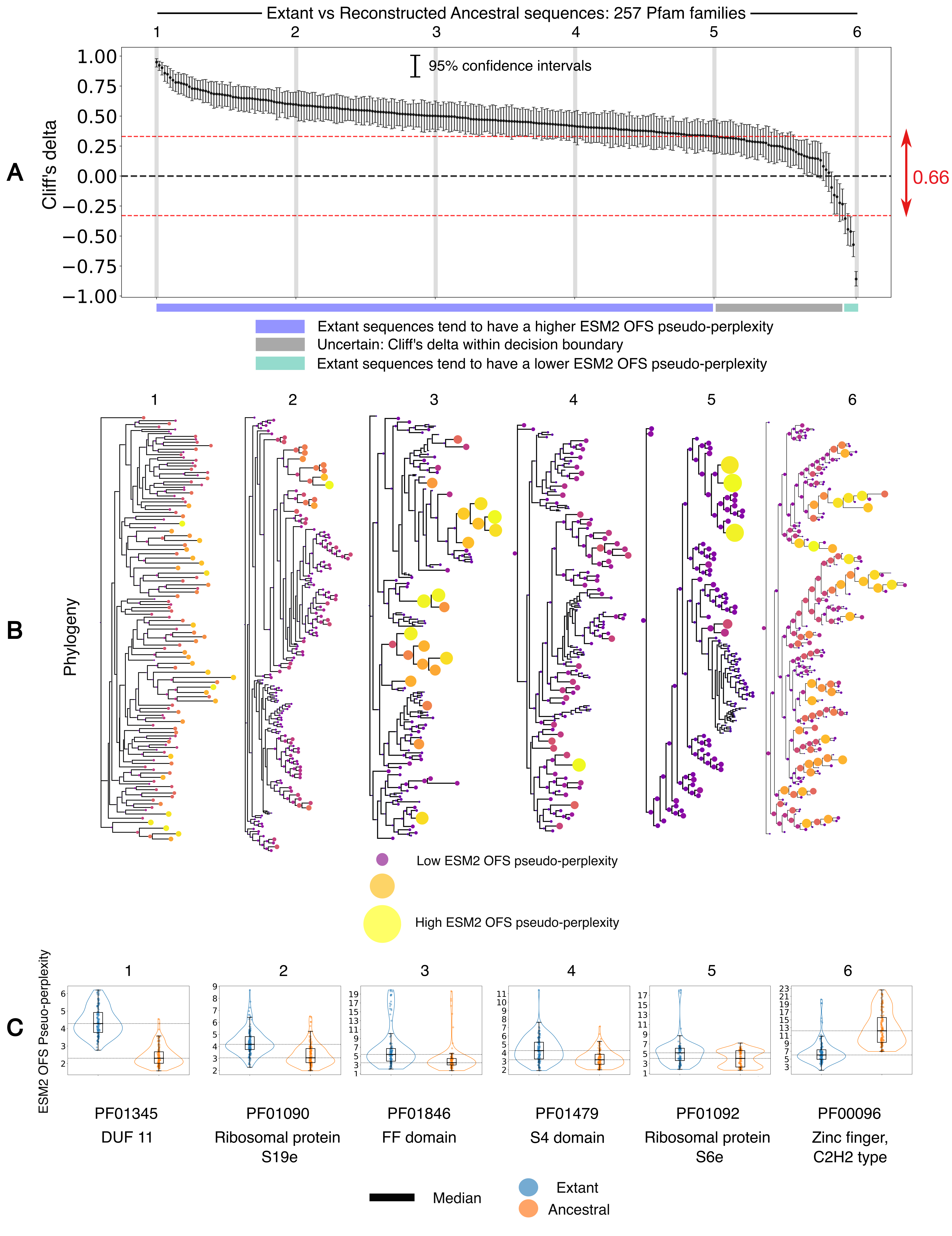}
    \caption{\small (A) A plot of Cliff's delta for the ESM2 OFS pseudo-perplexity between extant                     and reconstructed ancestral sequences for 257 protein families. The families                       are arranged in descending order of the Cliff's delta between the two sets.                        The gray vertical bars are used to demarcate the families shown in detail in                       B and C. The dotted line in red denotes the decision boundary. 
             (B) Phylogeny of the respective families that correspond to the gray bars in A. The interior nodes (reconstructed ancestral sequences) in the tree tend to have lower ESM2 OFS pseudo-perplexity values for the 5 trees that have a positive Cliff's delta, while the leaf nodes (extant sequences) tend to have a lower ESM2 OFS pseudo-perplexity in the last tree where the Cliff's delta value is negative. The size and color \cite{huerta2016ete, cock2009biopython} of the pseudo-perplexity markers (solid circles) is only meant to be compared within a tree, and not across different trees.
             (C) The distribution of the fitness scores in the two sets of sequences visualized in a box plot for the six families marked by the gray bars in A.}
    \label{fig:figure_3}
\end{figure}

An alternative line of reasoning for the phenomenon asserts that the inferred ancestral sequences appear to possess superior functional properties because the true ancestral sequences did in fact possess those properties. The functional properties of a protein are a product of the environment that it operates in. So, only proteins with a superior stability could have been functional in a pre-Cambrian hot bath, or a higher oxidative stress environment, or the higher radiation levels of a more tumultuous earth of the past~\citep{gaucher2008palaeotemperature, trudeau2016potential}. Furthermore, there have been proposals where sub-functionalization over evolutionary time-scales can mediate a generalist-to-specialist transition, wherein promiscuous ancestral proteins progressively become more specialized in function~\citep{risso2013hyperstability, risso2014thermostable, lynch2000probability, wheeler2016thermostability}. This transition can also be used to argue for systematic differences between the two sets, however a consensus for the existence of such a trend has also not been established \citep{wheeler2016thermostability}. 

Disentangling the confounding variables in this setup has not been straightforward. Regardless of the underlying reason, we set out to investigate if we can detect the disparity in function between extant and reconstructed ancestral sequences. Addressing this problem entails building a framework to distinguish between the statistical behavior of populations of sequences. We use ESM2 OFS pseudo-perplexity as a scalar proxy for function to assign the fitness scores for individual sequences. We then use Cliff's delta, a non-parametric estimator of effect size to compare the fitness scores of the respective populations \cite{cliff1993dominance}. Cliff's delta measures the tendency of elements in $A$ to be greater than those in $B$ without making any assumptions about the distributions that the values are drawn from.  For $A = \{a_1, a_2, \ldots, a_m\}$ and $B = \{b_1, b_2, \ldots, b_n\}$, where the elements correspond to the fitness scores for the respective sequences in the two sets in this setting, Cliff's delta ($\delta$) is given by:

\begin{equation}
\delta(A,B) = \frac{\sum_{i=1}^{m}\sum_{j=1}^{n} \text{sgn}(a_i - b_j)}{mn}
\end{equation}
where, $sgn$ denotes the signum function, which outputs ${1, 0, -1}$ based on the input's sign. Cliff's delta between the fitness scores of the two sets of sequences can thus be used as a measure for the disparity in function between the populations. 

We calibrated this metaphorical instrument using data from ProteinGym DMS assays and found that the consistency in the sign of the inferred and true value of Cliff's delta strongly depends on the magnitude of the inferred value: the larger the magnitude, the more likely it is that the sign is inferred correctly. The true value in this context is defined to be the negative Cliff's delta between the experimentally measured fitness scores of the sequences in the two sets. The negative sign is introduced in the expression to match its sign with the inferred disparity that uses the ESM2 OFS pseudo-perplexity of the sequences in the calculation. This observation prompted us to use a symmetric decision boundary of 0.33 around 0 to reliably ascertain the direction of the discrepancy in function. In this scheme, values that lie within the boundary are discarded as error-prone, and therefore uncertain, while values that lie outside of it can be used as confident estimates for the sign of the disparity in function (see Appendix:Decision Boundary for more details).

With a framework in place, we can return to addressing the original question: Can we capture the widely-reported elevated stability of reconstructed ancestral sequences? Towards this end, we procured sequence alignments for 257 protein families from the Interpro database \citep{paysan2023interpro}. The pfam families \cite{mistry2021pfam} used in this analysis were not chosen randomly, however we were cognizant so as not to introduce a systematic bias that would influence the outcome of the experiment (see Appendix:Ancestral state reconstruction for details). We used IQTREE2 \cite{minh2020new, kalyaanamoorthy2017modelfinder} to infer the phylogeny of the families and FastML \cite{ashkenazy2012fastml} is used for ancestral state reconstruction. A positive value for Cliff's delta in this setting indicates that the extant sequences tend to have a higher ESM2 OFS pseudo-perplexity compared to the reconstructed sequences, while a negative value indicates the opposite trend. We found that Cliff's delta has a positive value greater than the chosen threshold of 0.33 for 205 out of 257 families (79.76\%), and a negative value less than -0.33 for 5 out of 257  families (1.94\%) as shown in Figure 3. We note that some of the families with the negative value of Cliff's delta are polyphyletic in origin, which may explain the higher pseudo-perplexity of the reconstructed ancestors. The remaining 47 families lie in the uncertain range where estimates for the sign of Cliff's delta are not reliable.

The results imply that reconstructed ancestral sequences are predicted to have a higher stability for the majority of protein families under consideration, as per the language model assessment. Thus, we are in fact able to discern the elevated stability of reconstructed ancestral sequences using the fitness measure. We note that we have not experimentally validated the claim regarding the enhanced stability for the specific protein families that we have evaluated in this analysis, and are relying on the empirically observed trend being sound in general. What is certainly true however, is that a lower ESM2 OFS pseudo-perplexity is predictive of a higher stability as per the performance of the fitness measure on DMS assays.

It is tempting to speculate whether a lower pseudo-perplexity in the reconstructed sequences implies that a consensus effect is at play in the reconstruction inference process, since pseudo-perplexity does relate to typicality. We refrain from indulging in this temptation, ever so sweet as it may be. Establishing and quantifying such a bias in the reconstruction procedure requires a more careful analysis that we will pursue in an upcoming venture. For now, we only submit that the model's assessment can be used to detect the known disparity in function between the two sets of sequences. Ascribing a reason for precisely why this is the case is not straightforward, but the fact that the language model is in tune with this trend is intriguing. The model has thus picked up on this design principle independently through self-supervised means, lending additional support to the idea that large language models are more than just stochastic parrots \citep{arora2023theory}. 

\section{Exploring Protein Morphospace}

Protein design constitutes a broad range of design goals such as enhancing the functional properties of existing sequences, modifying known sequences to acquire new functions, or even generating altogether novel sequences with as of yet unseen folds \citep{notin2024machine}. The common thread in the endeavor is to wade through protein morphospace and sample sequences that best satisfy some user-defined constraints. Machine learning models have been used to navigate through this vast space by pruning it down to the much smaller set of presumably viable states- states that have a high likelihood as per the model's assessment. Several such deep learning powered workflows have emerged that have been successfully deployed to generate functional protein sequences \citep{verkuil2022language, hie2022high, sumida2024improving, nijkamp2023progen2}. 

\begin{figure}[htbp]
    \centering
    \includegraphics[width=0.8\textwidth]{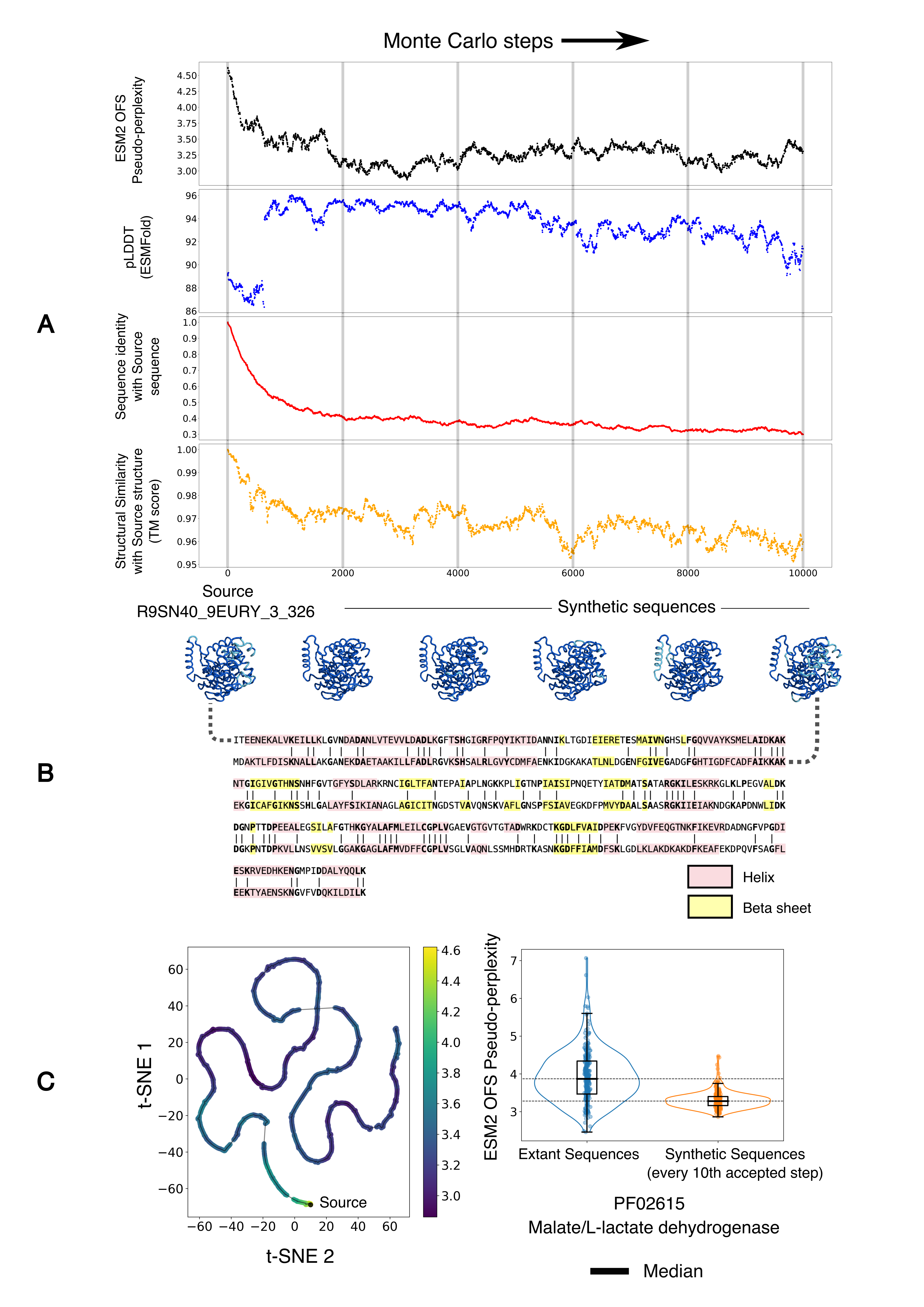}
    \caption{\small (A) A low temperature Monte Carlo run with sequence profile-based mutation                          proposals and ESM2 OFS pseudo-perplexity as the energy function. The folding                     confidence of the sequences is consistently high, as measured by the pLDDT                         score from ESMFold, and loosely tracks the negative energy. The sequence                           identity with the source sequence also decreases over the course of the                            procedure, however the fold similarity as measured by the TM score remains                         high.
             (B) A visualization of the structures assumed by the intermediate sequences, folded via ESMFold, along with the sequence alignment between the source sequence and the final sequence.
             (C) Left: A visualization of the path traced out by the intermediate sequences through a t-SNE projection of the mean-pooled ESM2 embeddings. The points are colored by their ESM2 OFS pseudo-perplexity. 
             Right: A visualization of the difference between the fitness scores of the two sets of sequences in a box plot.}

    \label{fig:figure_4}
\end{figure}

The design goals in such generative tasks can be broadly framed as de novo sequence generation, family-specific sequence generation or structure-conditioned generation. Decoding language models are particularly well-suited for the de novo sequence generation task. These models utilize an autoregressive sampling scheme where the sequence manifests one residue at a time, and can be used to sample sequences that are far from natural proteins in sequence identity. Family-specific sequence generation can be performed by fine-tuning the decoding model on sequences drawn from the family of interest \citep{nijkamp2023progen2, shuai2021generative}. Alternatively, hybrid language models offer a fine-tuning-free way to generate family-specific sequences by using the sequences from the family to condition the generation process \cite{truong2024poet, sgarbossa2023generative}. Likewise, inverse folding models have enabled structure-oriented sequence generation as they can model the distribution of protein sequences that acquire some given input structure \cite{hsu2022learning, dauparas2022robust}. 

In contrast to the above generative schemes, an iterative refinement workflow involves starting with a seed sequence that is progressively morphed to better fit the provided design constraints. Such workflows typically employ a Markov chain Monte Carlo framework for the refinement procedure. The design objective is framed as an energy minimization problem, where the energy function encodes the constraints that a successful design needs to satisfy: sequences that best satisfy the constraints have the lowest energy. The protocol entails sampling mutation proposals from some designated sampling distribution that can be accepted or rejected based on the energy of the resultant sequence. The optimization involves making such proposals in an iterative fashion to improve the design over time, yielding low energy sequences that satisfy the provided constraints. The generated pool of sequences can be scored to identify the most promising candidates, and the selected candidate sequences can then be experimentally evaluated for downstream use \cite{biswas2021low, hie2022high, verkuil2022language, anishchenko2021novo}.

We contend that the one fell swoop technique combined with an iterative refinement workflow is particularly well suited for the family specific sequence generation problem. The mutation proposals used in these workflows are typically sampled uniformly over the single residue mutational landscape of the sequence. We propose to use the sequence profile obtained from the ESM2 OFS setup as the sampling distribution for the proposals instead, as it is poised to automatically upsample sensible mutations and therefore increase the exploration efficiency of the procedure. A combination of sequence likelihoods drawn from models trained on different data modalities and modeling objectives could be used to define a custom energy function to generate high fitness variants of the input seed sequence. Such an ensembled setup can potentially lead to higher downstream success by serving as a powerful prefilter for viability and enhanced function in the sequence generation process \cite{johnson2024computational}. 

As a proof of concept, we use ESM2 OFS pseudo-perplexity as the energy function to generate variants of the Malate/L-lactate dehydrogenase family of enzymes that are predicted to have an enhanced stability and function via low temperature sampling (Figure 4). The results show that the energy of the generated sequences equilibrates to a lower range of values over the course of the run, while the folding confidence of the sequences as measured by the pLDDT score (ESMFold) remains high (Figure 4:A). In addition, the generated sequences progressively decrease in their shared sequence identity with the source sequence, while the structural similarity with the initial structure as measured by the TM score remains high (Figure 4:A). The procedure can thus be used to generate a diverse set of candidate sequences that are predicted to have enhanced properties by enabling explicit control over the fitness of the generated variants through the sampling temperature.

\section{Discussion}

Here we introduce One Fell Swoop, a distillation-inspired trick~\citep{hinton2015distilling} for fast single-pass estimation of pseudo-perplexity for protein sequences. Distillation involves training a compact model to replicate the output of a larger, more complex model, by using the nuanced assessments of the larger model as the target output for the compact model. In a similar vein, OFS entails training a small projection model to generate the masked profile of a protein sequence, as predicted by the much larger encoding language model. The technique deviates from traditional distillation in allowing the projection model to access the encoder’s unmasked residue embeddings to enable accurate and parallelized estimation of the masked profile. The estimated masked profile serves as a valuable intermediate in the pseudo-perplexity calculation as it can be used to identify sub-optimally occupied positions in a sequence, and pinpoint the options that fit best in the given context. 

This approach circumvents the need to fine-tune the encoder specifically for pseudo-log-likelihood estimation, as has been proposed in the natural language setting~\cite{salazar2019masked}. Instead, the specialized OFS model handles the regression task by utilizing the full resolution of the sequence representation from the encoder. This representation consists of the set of all residue embeddings that are generated for each residue within the sequence. This is in contrast to the typical use of a singular coarse-grained representation of the sequence obtained by averaging over the residue embeddings, or the use of the classification token embedding as is done in BERT~\cite{devlin2018bert}, for sequence-level regression problems of this sort~\cite{elnaggar2021prottrans, salazar2019masked}. We are leveraging the small size of the vocabulary for protein sequences in making this design choice, as it likely makes the regression task easier.

We used ESM2 (650M) to demonstrate the efficacy of OFS pseudo-perplexity as a fitness estimation metric by benchmarking its performance on the ProteinGym dataset. We show that the scoring scheme enables encoding models to evaluate indels, and more notably, we find that ESM2 OFS pseudo-perplexity defines a new state of the art on the ProteinGym indels benchmark. The metric’s strong performance at fitness estimation for stability-based assays prompted us to investigate its potential to detect the elevated stability reported in reconstructed ancestral sequences. We found that reconstructed ancestral sequences tend to have a lower ESM2 OFS pseudo-perplexity relative to extant sequences for the majority of the protein families that we considered. 

Reconstructed ancestral sequences thus constitute a class of synthetic sequences that by and large score higher on predicted fitness than natural extant sequences- a trend that is consistent with empirical measurements of protein function. The analysis thus reveals that ESM2 has learned this non-obvious protein design principle on its own in an organic self-supervised manner. The design principle is non-obvious not just because the underlying mechanism remains mysterious, but also because it requires the model to assign higher fitness scores to reconstructed sequences when it has only been trained on extant sequences~\cite{shaw2023removing}. This feat, much like the model’s performance on the fitness estimation benchmark, is a testament to its ability to generalize well beyond what it has seen during training. 

We also propose a Markov chain Monte Carlo based iterative refinement workflow for protein design where the sequence profile is used as the proposal distribution, and ESM2 OFS pseudo-perplexity is used as the energy function. The temperature parameter for the process can be tuned to selectively sample sequences that are predicted to have enhanced functional properties. We have observed that short runs reliably generate sequences with a high folding confidence as measured by their pLDDT scores. However, we also found that the folding confidence of the generated sequences eventually degrades over the course of a single very long run. This phenomenon can in part be traced to an effect that arises due to in-context learning, where pseudo-perplexity can remain low even when the sequence being evaluated is nonsensical. The emergence of imperfect repeated motifs within the sequence over the course of an extended run is a hallmark of this effect. In addition, these models have been shown to suffer from phylogenetic biases that stem from uneven species representation in the training dataset. These biases can lead the generated sequences to drift towards favored clades, which can in turn lead to suboptimal design outcomes when working with under-represented sequences~\citep{ding2024protein}. Therefore, these factors should be carefully considered when using these generative methods. Nevertheless, the model’s ability to uncover non-obvious design principles bolsters the case for its use as a vehicle to explore protein morphospace. 

\section{Acknowledgements}
We thank Michael Abbott, Jose Betancourt and Casey Dunn for close reads and useful feedback on the manuscript.  This work was supported by NIH R35GM138341 (BBM). We also thank the Yale Center for Research Computing, ITS Cloud, and AWS for guidance and use of AWS Cloud computing infrastructure through Cloud Credits for Research Program.  

\bibliographystyle{unsrt}
\bibliography{pseudo_perp}

\section{Appendix}

\subsection{Pseudo-perplexity in One fell swoop}
The one fell swoop setup is composed of an ensemble of 8 multi-layer perceptron models that use ReLU activations and layer normalization. The training data is generated by filtering the Swissprot database to sequences shorter than 700 residues and clustering the remaining sequences at 50\% identity via mmseqs2. The representative sequences for the clusters are then used for the sequence profile calculation that is performed by masking the residues one at a time. The probability vector for each position is calculated via the softmax operation over the logits for the 20 natural amino acid residues. Sequences that share greater than 50\% identity with reference sequences in the ProteinGym dataset were withheld from being used in training. We use a 99 to 1 training-validation split, and the model is trained to minimize the cross-entropy loss between its prediction and the output from the one-at-a-time calculation via AdamW. 

\subsection{Evaluating OFS pseudo-perplexity as a fitness estimator}
Pseudo-perplexity via the one-at-a-time calculation is calculated for a subset of 61 DMS assays in ProteinGym, where the sum of the lengths of the variants in the assay is less than the chosen threshold of 40,000. The masked marginal heuristic is calculated by masking all mutated positions at the same time, and the probability vector for each position is calculated via the softmax operation over the logits of the 20 natural amino acids. The aggregate mean in Table 1 and Table 2 is calculated by aggregating over the uniprot id, and then aggregating over all the assays.

\subsection{Ancestral state reconstruction: Old is Gold?}

\textbf{Decision Boundary} ESM2 OFS pseudo-perplexity used in tandem with Cliff's delta, allows us to measure the difference in fitness between two populations of sequences. This metaphorical instrument has to be calibrated, before it can be put to use. This involves characterizing the accuracy of its predictions on data where the ground truth is known, and thereby evaluating its reliability and effectiveness at the task. We generate the evaluation data for this purpose by partitioning the variants in a single DMS assay into two randomly drawn populations. The sampling scheme to partition the sets is tuned by a bias parameter that biases the procedure to selectively sample high or low fitness variants within a given DMS assay. Each partition contains 200 sequences, unless the total number of variants in the assay is fewer than 400. For these cases, we use all the variants in the assay to generate the partitions. Several such partitions are created for all of the 283 assays in ProteinGym yielding diverse sets of populations for which the true value of the disparity in function can be calculated. The true value in this context is defined to be the negative Cliff's delta between the experimentally measured fitness scores of the sequences in the two sets. The negative sign is introduced in the expression to match its sign with the inferred disparity that uses the ESM2 OFS pseudo-perplexity of the sequences in the calculation. The reliability of the reading is assessed by the consistency in the sign of the inferred and true Cliff's delta. This evaluation metric prioritizes getting the direction of the disparity right, even if the degree of the disparity in function between the two sets is not inferred accurately.

The results indicate that the consistency in the sign of the inferred and true value, strongly depends on the magnitude of the inferred value: the larger the magnitude, the more likely it is that the sign is inferred correctly (Figure 5:A). The plots in figure 5:A correspond to a bias parameter of 1 (left), 0.5 (middle) and mixed (right), where the parameter is itself a random number in the mixed case. The plots are generated by creating 1000 sets for each of the 283 assays and the calculation for the true and inferred Cliff's delta is performed by using the DMS fitness scores and ESM2 OFS pseudo-perplexity respectively. This allows us to define the range of values of the inferred Cliff's delta that can be used as reliable predictors for the direction of the disparity in function between the two sets. Such a range can take the form of a decision boundary: values that lie within the boundary are discarded as error-prone, and therefore uncertain, while values that lie outside of it can be used as confident estimates. The continuously varying estimate of the disparity in function between sets A and B, measured via Cliff's delta, is thus mapped to three discrete states in this setup: (i) values in A tend to be greater than values in B,  (ii) values in A tend to be less than values in B, and (iii) values in the two sets cannot be reliably distinguished by the procedure. 

The efficacy of the classification method is predicated on the placement of the boundaries that define the three states. Choosing the precise value for the threshold is a compromise between how much of the true signal we are willing to forego, and how many errors we are willing to tolerate: too stringent of a value for the threshold will inevitably disregard a significant proportion of cases where the discrepancy in function is truly present, whereas a low cutoff will allow errors to creep in. We choose to err towards the side of caution, and opt for a symmetric threshold of 0.33 around 0. This threshold allows for a high prediction accuracy for the direction of the discrepancy in function, which in turn instills confidence in the method's use as a predictive tool. The higher accuracy comes at the expense of dismissing the readings for a sizable chunk of the pairs in the evaluation sets as unreliable. 

We simulate a test run of the procedure on specially constructed populations of sequences for each of the 283 assays: minimal disparity and maximal disparity sets (Figure 5:B). The minimal disparity sets are created by rank ordering the variants in ProteinGym DMS assays by their experimentally measured fitness scores and then partitioning them by selecting sequences alternately, while the maximal disparity sets are created by partitioning the variants into the upper and lower halves of the fitness spectrum. For assays where the total number of variants is greater than 400, only the top and bottom 200 sequences arranged by the fitness scores are used in this analysis, while all the sequences are used for assays with fewer than 400 variants. The 95\% confidence intervals are generated via sampling with repetition. The primary point of this exercise is to visually inspect the inferred Cliff's delta curves for datasets with a low and high degree of disparity in function. We see that all of the minimal disparity sets end up with Cliff's delta values that lie inside the decision boundary (Figure 5:C). This indicates that the detected disparity is too low for its direction to be reliably inferred by the procedure. For the maximal disparity sets we see that the procedure correctly ascertains the direction of the discrepancy in function for a significant proportion of pairs 243 out of 283 (85.86\%), however the remaining 40 pairs do end up in the gray zone of uncertainty (Figure 5:D). \newline

\begin{figure}[htbp]
    \centering
    \includegraphics[width=0.85\textwidth]{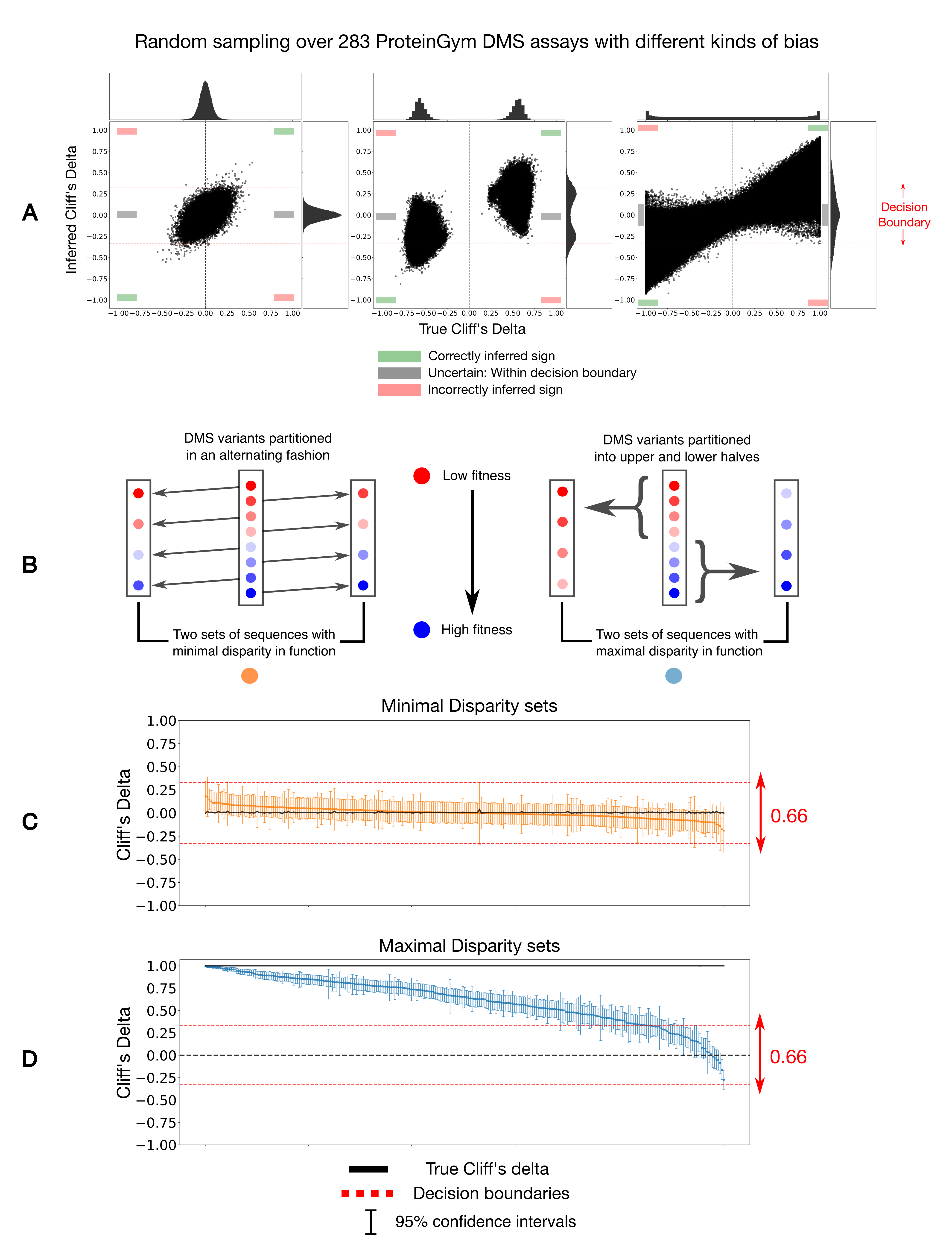}
    \caption{\small (A) Variants in the DMS assays are partitioned into two sets of sequences via random               sampling to calculate the true and inferred value of Cliff's delta for the populations.            Bias is introduced in the sampling process to generate populations with a larger                   disparity in function. Inferred values that are larger than 0.33 in magnitude are largely          inferred correctly.
            (B) Construction process for the minimal and maximal disparity sets: minimal disparity sets are partitioned into two populations via alternate sampling while maximal disparity sets are partitioned into upper and lower halves.
            (C) The Cliff's delta curve for the minimal disparity sets lies within the decision boundary indicating that the measured disparity between the sets is too small for its direction to be reliably inferred.
            (D) The Cliff's delta curve for maximal disparity sets is positive for a significant proportion of the DMS assays.}
    \label{fig:figure_5}
\end{figure}

\noindent \textbf{Ancestral state reconstruction}
Seed alignments for pfam families with codes from PF00001 to PF03000 were procured from the Interpro database. Only families with a sequence length (including gaps) less than or equal to 500, at least 100 unique sequences and at most 250 total sequences, a gap content of less than 50\%, and no non-standard amino acids were considered for the analysis. These filters yield 301 protein families in total. We used IQTREE2 to infer the phylogeny of the families using the MFP model finder flag. The inferred tree is fed to FastML for ancestral state reconstruction, and the single most likely sequence for each node in the tree is used in the analysis. We ran the outlined procedure for all 301 families, however FastML jobs only finished for 257 families even after 96 hour long runtimes for the individual jobs. The results for these 257 families are presented in the analysis.

\subsection{Exploring protein morphospace}
A state-dependent proposal distribution necessitates a modification in the acceptance probability for the regular Metropolis criterion used in symmetric proposal schemes. We denote the state of the sequence at time t by $X_{t}$. Let u be a mutation that changes the sequence $X_{t}$ to $X'$ and v be the inverse mutation that changes sequence $X'$ to $X_{t}$, then the probability of accepting the mutation u for the sequence $X_t$ is given by

\[
A(u|X_{t}) = \text{min}(1, \frac{P(X')}{P(X_{t})} \frac{P(v|X')}{P(u|X_{t})})
\]

Here P(X) is given by $e^{-E(X)/T}$, where $E$ is the energy function and $T$ is the temperature. The probabilities for $u$ and $v$ can be calculated using the sequence profiles of the respective states. We use ESM2 OFS pseudo-perplexity as the energy function in Figure 4, and use a temperature of 0.01 for sampling the sequences. Note that only substitutions are used as mutation proposals in Figure 4.

\end{document}